\documentclass[showpacs,aps,amsmath,letterpaper,nofootinbib,reprint,showkeys,superscriptaddress,longbibliography,pra]{revtex4-2}
\usepackage{dcolumn}    
\usepackage{bm}         
\usepackage{ifpdf}
\usepackage{amssymb,lineno,amsfonts}
\usepackage{graphicx}   
\usepackage{bbm}
\usepackage{mathrsfs}
\usepackage{mathtools}
\usepackage{upgreek}

\usepackage{epstopdf}
\usepackage{setspace}
\usepackage{hyperref}
\usepackage{cleveref}
\usepackage{bbold}
\usepackage[matrix,frame,arrow]{xypic}
\usepackage{float}
\usepackage{soul}
\usepackage{tikz}
\usepackage{natbib}
\usepackage{color} 
\usepackage{subfigure}
\usepackage{bbold}
\usepackage{tikz}
\usepackage{amsthm}
\usepackage{academicons}
\usepackage[english]{babel}

\definecolor{med-blue}{RGB}{25,25,112} 
\hypersetup{colorlinks, linkcolor={blue},citecolor={blue}, urlcolor={blue}}

\newcommand{\abs}[1]{\vert{#1}\vert}

\usepackage{orcidlink}

\begin{document}
    \title{Quantum Sensing via Large Spin-Clusters in Solid-State NMR: \\ 
    Optimal coherence order for practical sensing}
	\author{C Alexander}
	\email{conan.alexander@students.iiserpune.ac.in}
	\author{T S Mahesh}
	\email{mahesh.ts@iiserpune.ac.in}
	
	\affiliation{Department of Physics and NMR Research Center,\\
		Indian Institute of Science Education and Research, Pune 411008, India}
	
\begin{abstract}
{Quantum entanglement has long been recognized as an important resource for quantum sensing. In this work, we demonstrate the use of multiple-quantum solid-state NMR for quantum sensing by creating, manipulating, and detecting large clusters of correlated nuclear spins. We show that such clusters can sensitively detect pulse-width jitters in radio-frequency control fields at the level of tens of nanoseconds.
By analyzing the response of high-order quantum coherences to these control-field jitters, we investigate the critical interplay between the enhanced sensitivity offered by large coherence orders, their relative distributions, and their varying susceptibility to decoherence. We further demonstrate that, even within a non-uniform distribution of coherence orders, there exists an optimal maximum coherence order that maximizes sensing efficiency.
To support our interpretation, we supplement the experimental results with a simplified numerical model that estimates the corresponding quantum Fisher information. These results support the solid-state NMR platform as a valuable testbed for investigating many-body quantum metrology protocols.
}
\end{abstract}

\maketitle	

\section{Introduction}
\label{sec:Introduction}

Quantum technologies have witnessed sustained growth over the past two decades, driven by the promise of advantages over classical approaches in computation, communication, and sensing. Among these, quantum sensing—or quantum parameter estimation—aims to enhance the measurement precision of physical quantities such as  electric or magnetic fields and has emerged as a central theme in both theoretical and experimental research \cite{quantumsensingpaola,montenegro2025quantum, Giovannetti2011}.

Theoretically, strategies utilizing quantum entanglement can achieve enhanced sensitivity in ideal conditions \cite{Lee2002,beatingSQL} with a large body of work exploring robust advantages even in the presence of noise \cite{entangledfockstates,nisqera,wu2021non,chin2012quantum,zhang2022non,berrada2013non,wu2020quantum,altherr2021quantum,mirkin2020quantum}. This promise has spurred rapid experimental progress across diverse physical platforms, including ion traps \cite{iontrap1,iontrap2,iontrap3}, nitrogen vacancy centers \cite{nvcenter1,nvcenter2,nvcenter3,nvcenter4}, superconducting circuits \cite{supercond1,supercond2,supercond3}, nuclear magnetic resonance (NMR) systems \cite{nmr5, nmr1, shukla2014noon, nmr4}, and many more. Among these, solid-state systems are particularly compelling due to their high spin densities, offering a potential path towards scalable quantum sensors, although many demonstrations have been limited to uncorrelated ensembles \cite{quantumsensingpaola,solidstatesensors}.  

Despite this progress, a significant gap persists between theoretical benchmarks and experimental realities \cite{montenegro2025quantum, Giovannetti2011}. The ultimate sensitivity is theoretically bounded by the quantum Fisher information (QFI), which requires optimizing over all possible measurements. However, implementing QFI-saturating optimal measurements is often experimentally prohibitive, especially in complex, interacting many-body systems. Furthermore, experimental imperfections and environmental noises diminish the realistically achievable precision, prompting the search for alternative methods of analysis \cite{luis2013alternative,montenegro2025quantum,rubio2018nonasymptoticmetrology}.



Quantum sensing or parameter estimation focuses on inferring the value of a quantity $\alpha$ from measurements on a state $\rho(\alpha)$ which encodes the quantity as a parameter. The uncertainty in estimating $\alpha$ as quantified by the standard deviation $\delta\alpha$ is bounded by the Cram\'er-Rao inequality \cite{Giovannetti2011},
\begin{equation}
\delta\alpha\geq 1/\sqrt{N_m\mathcal{F}_c}.   
\end{equation}
Here $N_m$ is the number of measurements and $\mathcal{F}_c$ is the classical Fisher information (CFI) given by,
\begin{equation}
\mathcal{F}_c = \sum_i p_i(\alpha)[\partial_\alpha \ln{p_i(\alpha)}]^2,    
\end{equation}
with $p_i$ being the probability of the $i^{th}$ measurement outcome and the sum running over all possible outcomes. Generalizing to the quantum case, we get the quantum Cram\'er-Rao inequality (qCR),
\begin{equation}
\delta\alpha\geq 1/\sqrt{N_m\mathcal{F}_q},    
\end{equation}
where $\mathcal{F}_q(\alpha)$ is  the optimum Fisher information $\mathcal{F}_c$ over all possible measurements. In the classical case, Fisher information at best scales linearly with system size $N$. This scaling is referred to as the standard quantum limit (SQL). In quantum systems, however, $\mathcal{F}_q$ can scale quadratically with the system size, i.e., $\mathcal{F}_q \propto N^2$. This scaling limit is called the Heisenberg limit (HL) \cite{Giovannetti2011}. 

One of the most common protocols in this field is phase estimation using Ramsey interferometry, in which quantum states with high coherence order are employed to estimate quantities by encoding them in their phase. In this type of sensing, the scaling advantage of larger $N$ is manifested through the ability to generate higher coherence orders, with the bound on precision inversely scaling with coherence order in the absence of noise \cite{Giovannetti2011}.  In practice, however, we may realize a range of coherence orders with different relative proportions.
Moreover, the larger $N$ or larger coherence order also suffers from stronger decoherence, which naturally reduces the sensing efficiency.  These issues effectively lead to a trade-off between the HL-allowed quantum advantage that grows with coherence order and the quantum memory that diminishes with the coherence order, further compounded by the nonuniform mixture of coherence orders.  Given this scenario, our motivation is (1) to study this interplay and determine the existence of an optimal coherence order, and (2) to find the efficiency of quantum sensing with such a spin cluster composing a nonuniform mixture of coherence orders.

In this study, we propose and explore the use of multiple quantum  solid-state NMR as a platform for quantum sensing. This platform is adept at generating large-scale correlated spin clusters, observed as high-order quantum coherences \cite{baum}. 
Although the generation of large coherence orders has long been demonstrated in solid-state NMR \cite{baum,Krojanski2004,suterdecoherence}, there are few studies of their application for quantum sensing.  Our objective is to utilize the many-body quantum states in solid-state NMR as sensitive probes and demonstrate them as a rich testbed for quantum metrology protocols. 

The paper is organized as follows: 
In Sec. \ref{sec:Cluster Growth}, we describe the process of generating and detecting high coherence order states. In Sec. \ref{sec: Jitter Sensing}, we describe the experimental sensing of fluctuations in the control field, specifically the RF pulse-width jitters.  In Sec. \ref{sec:numerical model}, we discuss a numerical model to  complement the experimental findings. Finally, in Sec. \ref{sec:Conclusions}, we summarize the results and conclude with discussions.

\begin{figure*}
         \centering
         \includegraphics[trim=0cm 0.5cm 0.2cm 0cm,clip=,width=\textwidth]{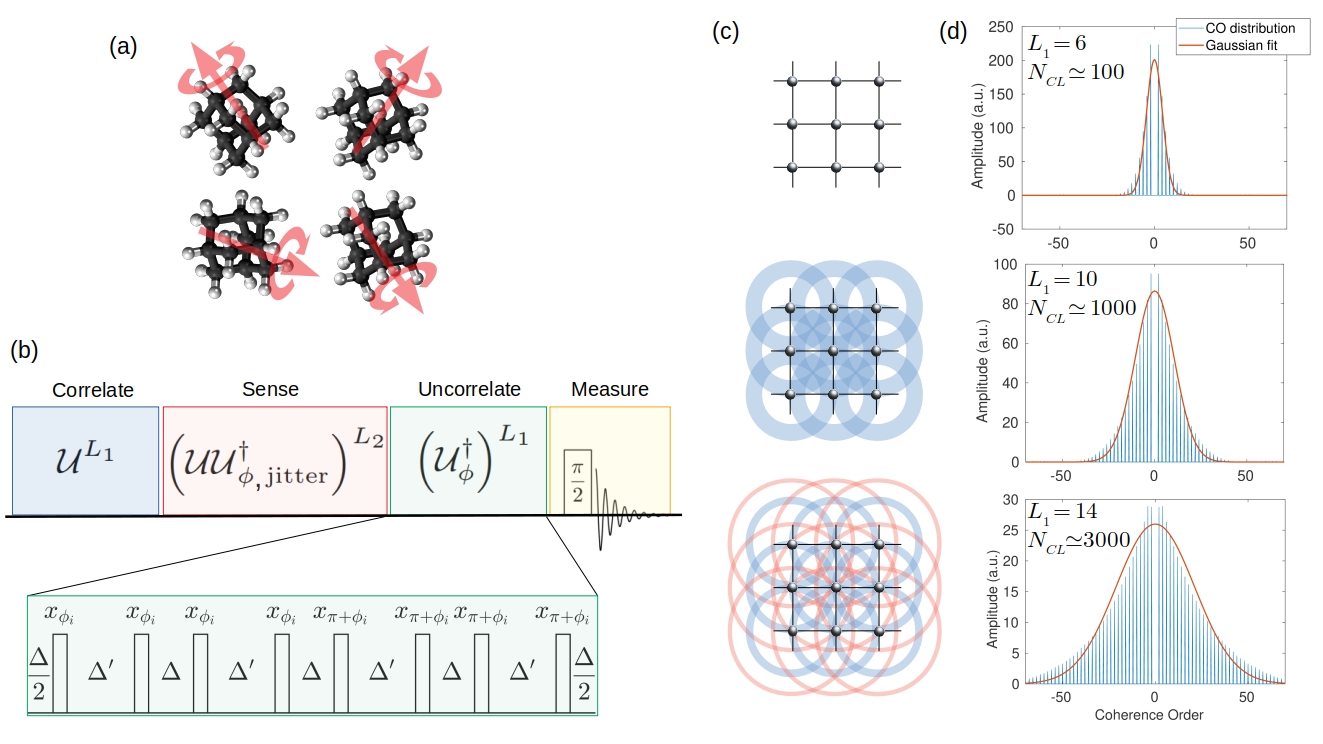}
         \caption{(a) Molecules undergoing random tumbling in adamantane plastic crystal.  (b) The overall sensing protocol and the RF sequence with eight $\pi/2$ pulses for generating or manipulating correlated spin clusters. The aim is to sense the RF pulse-width jitters introduced in $ \mathcal{U}_\phi^\dagger $ in the middle block with loop number $L_2$.  We set $\Delta = 1.5 ~\mu s$, $\Delta' = 2\Delta + \tau_{\pi/2}$, with $\tau_{\pi/2} = 2.9~ \mu s$, and vary the phase $\phi$ in the range $[0,\pi]$ in 181 steps.  (c) Illustrating the growth of correlated spin clusters as evolution under ${\cal H}_\mathrm{eff}$ progresses with increasing $L_1$ from top to bottom. (d) The experimentally obtained distributions of coherence orders at different values of $L_1$ (with $L_2=0$) and their Gaussian fits to extract the corresponding cluster sizes $N_{CL}$.  Here, the zeroth-order coherence, which is prone to spurious contributions, is suppressed by subtracting the average of signals before the Fourier transform. For $L_2>0$ implementing jitters, distortion in the coherence order distribution can be quantified by the distortion variance defined in Eq. \ref{eq:dv}.
    \label{fig:pulse-sequence}}
\end{figure*}

\section{Generating and detecting high coherence-order spin clusters}
\label{sec:Cluster Growth}
We choose $^1$H nuclear spins of adamantane (C$_{10}$H$_{16}$) a polycrystalline powder. All the experiments are carried out on a Bruker 500 MHz NMR spectrometer at a sample temperature of 300 K. Since adamantane is a plastic crystal at room temperature, molecules in  each crystallite exhibit translational order but randomly spin without any orientational order (see Fig. \ref{fig:pulse-sequence} (a)).  
The $^1$H nuclear spin qubits are initially in the uncorrelated thermal state, with the traceless part of the density matrix in the form
\begin{equation}
\rho_\mathrm{th} = \gamma \sum_i I^z_i,
\end{equation}
where $\gamma$ is the gyromagnetic ratio and $I^z_i$ is the z-component of the spin angular momentum of  $i$th spin.
The random molecular rotations average out the intramolecular dipolar interactions, while the intermolecular dipolar interactions are partially scaled down. This property of adamantane or similar plastic crystalline systems has made them popular choices for several quantum control experiments, such as Refs. \cite{baum,suterdecoherence,wei2020perturbation,kusumoto2021experimental}.
The system's natural evolution is governed by the residual intermolecular dipole-dipole interaction expressed by the Hamiltonian,
\begin{equation}
  H_{dd} = \sum_{i\neq j} d_{ij}( 3I^z_iI^z_j-\mathbf{I}_i \cdot \mathbf{I}_j ),
\end{equation}
where $d_{ij}$ is the residual dipolar coupling constant.
However, using an 8-pulse RF sequence \cite{baum} (see Fig. \ref{fig:pulse-sequence} (b)), the system can be made to evolve under an effective double-quantum Hamiltonian,
\begin{equation}
\mathcal{H}_{\text{eff}}=-\dfrac{1}{2}\sum_{i\neq j} d_{ij} (I^+_iI^+_j+I^-_iI^-_j).  
\end{equation}
As the spins evolve under this Hamiltonian, they develop larger and larger even-quantum coherences with neighboring spins, and thus the coherence order of states as well as the size of the quantum correlated cluster grows \cite{baum,Krojanski2004,suterdecoherence} (see Fig. \ref{fig:pulse-sequence} (c)).  
Our objective is to harness the large quantum correlated clusters for quantum sensing.

 To obtain the distribution of coherence order, an inverted and phase-shifted pulse sequence is applied that implements the net effective evolution,  
 \begin{equation}
  \mathcal{U}_{\phi}^{\dagger}(t) = e^{- i I^z \phi}e^{i \mathcal{H}_{\text{eff}}t}e^{ i I^z \phi},     
 \end{equation}
 for the same duration as the correlating duration, and a signal is obtained with a final detection $\pi/2$ pulse after a 3 ms delay that dephases any spurious transverse magnetization. A set of signals is obtained by varying $\phi$ in the range $[0,\pi]$ and the Fourier transform of the signals yields the coherence-order distribution as shown in Fig. \ref{fig:pulse-sequence} (d). 
 We can estimate cluster size $N_{CL}$ by fitting the coherence-order distribution to a Gaussian function, 
 \begin{equation}
  N_{CL} \sim \sigma^2/(4\ln 2),
 \end{equation}
 where $\sigma$ is the full width at half height (FWHH) of the Gaussian profile \cite{suterdecoherence}.  As shown in Fig. \ref{fig:pulse-sequence} (d), quantum correlated clusters of hundreds or even thousands of qubits are easily prepared with sufficient evolution loops $L_1$. 
 
\section{Sensing pulse-width jitter}
\label{sec: Jitter Sensing}
Although jitters can be introduced in any of the control parameters, such as pulse phase, pulse delays, and RF offset frequency, in this work we focus on jitters in pulse width.  We introduce pulse-width jitter of amplitude $\delta$ as random deviations from the nominal pulse parameters sampled from the uniform distribution over an interval of $[\tau_{\pi/2}-\delta/2,\tau_{\pi/2}+\delta/2]$, where $\tau_{\pi/2}$ is the duration of the nominal 90 degree pulse used in the 8-pulse RF sequence shown in Fig. \ref{fig:pulse-sequence} (b).   Since jitter is randomly applied, and it differs in each realization of the detection experiments, the effective evolution will now have a decohering component. The effective Hamiltonian $\mathcal{H}_{\text{eff}}$ can be replaced with an effective evolution superoperator,
 \begin{equation}
 \label{eqn:master}
     \mathcal{L}(\rho) = -i[\mathcal{H}_{\text{eff}},\rho] +  c \delta^2 \tau [V[V,\rho]]
 \end{equation}
 where 
 \begin{align}
     V = \frac{3}{2}\sum\limits_{i,j} d'_{ij} \hspace{0.1cm} (I^{y}_iI^{z}_j+I^{z}_iI^{y}_j),
 \end{align}
 and $c$ is a numerical constant (see Appendix A). 
 The resulting nonunitary evolution alters the intensities of various coherence orders, distorting their distribution.  We quantify  deviation of the jittered distribution $s_\delta(n_c)$ from the jitter-free distribution $s_0(n_c)$  up to coherence order $m_c$ by the  distortion variance,
 \begin{equation}
 D(\delta,m_c) = \dfrac{1}{m_c/2+1} \sum_{n_c=0}^{m_c} (s_0(n_c)-s_\delta(n_c))^2.
 \label{eq:dv}
 \end{equation}

\begin{figure*}
    \centering
     \includegraphics[trim=0cm 0.3cm 0cm 0cm,clip=,width=0.8\linewidth]{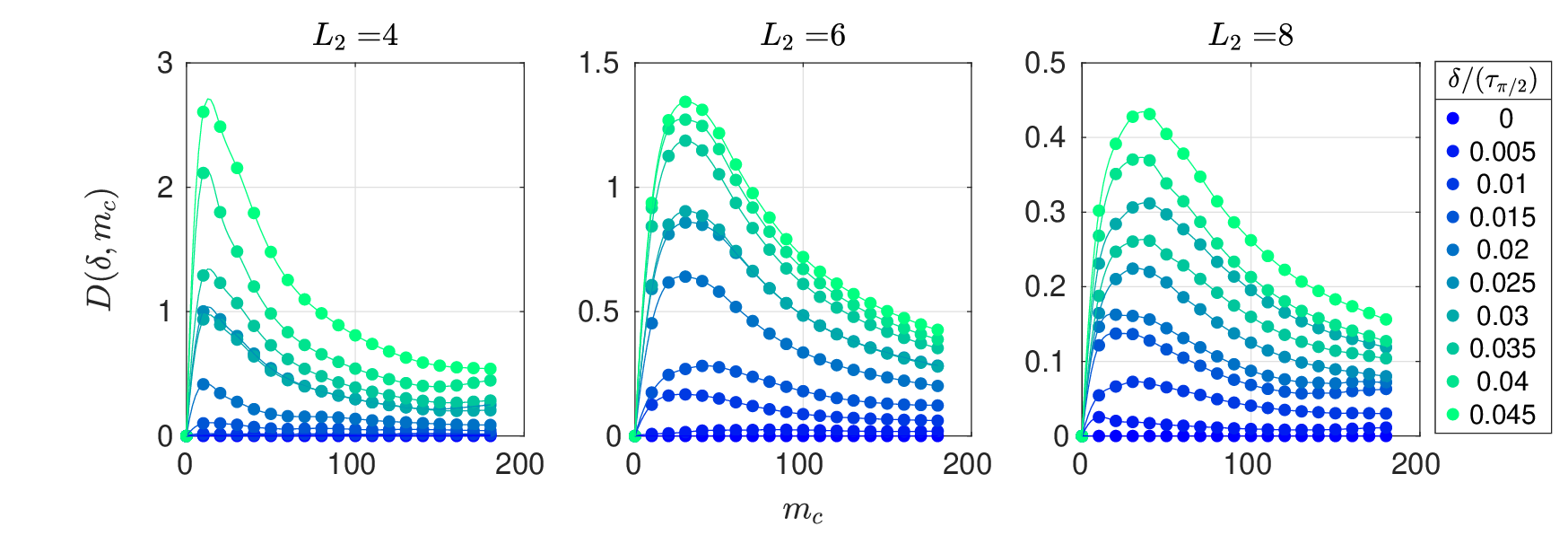}
    \caption{The dots are the experimental distortion variance $D(\delta,m_c)$ plotted versus the maximum coherence order $m_c$, for $L_1=10$, at different values of $L_2$, and with various values of the pulse-width jitter amplitudes as indicated in the legend bar. The lines are to guide the eye.
    \label{fig:cumulative} }
\end{figure*}

First, we analyze the contribution of clusters with a range of coherence orders to the sensing efficiency as measured by the  distortion variance. In the spirit of Heisenberg scaling, one may naively expect that the larger the maximum coherence order of a cluster, the higher will be the sensitivity.
Fig. \ref{fig:cumulative} shows the experimentally measured distortion variances $D(\delta,m_c)$  versus $m_c$ for $L_1=10$, and for various values of $L_2$ as well as $\delta$.
For lower maximum coherence orders, the distortion variance $D(\delta,m_c)$ indeed increases with $m_c$. However, in all the cases, $D(\delta,m_c)$ reaches a maximum at a certain $m_c$ value and then starts tapering. This behavior is due to lower proportions as well as  shorter lifetimes of higher coherence orders, which limits their sensing capacity, thereby exhibiting an optimum coherence order.

Now, given the large clusters of quantum correlated spins as our probe state, performing Fisher-saturating measurements may not be immediately feasible. Nevertheless, the distortion variance offers a way to directly analyze the coherence order distribution for sensing jitter amplitude $\delta$.


\begin{figure}[b]
    \centering
     \includegraphics[trim=1.2cm 0cm 2.2cm 0cm,clip=,width=\linewidth]{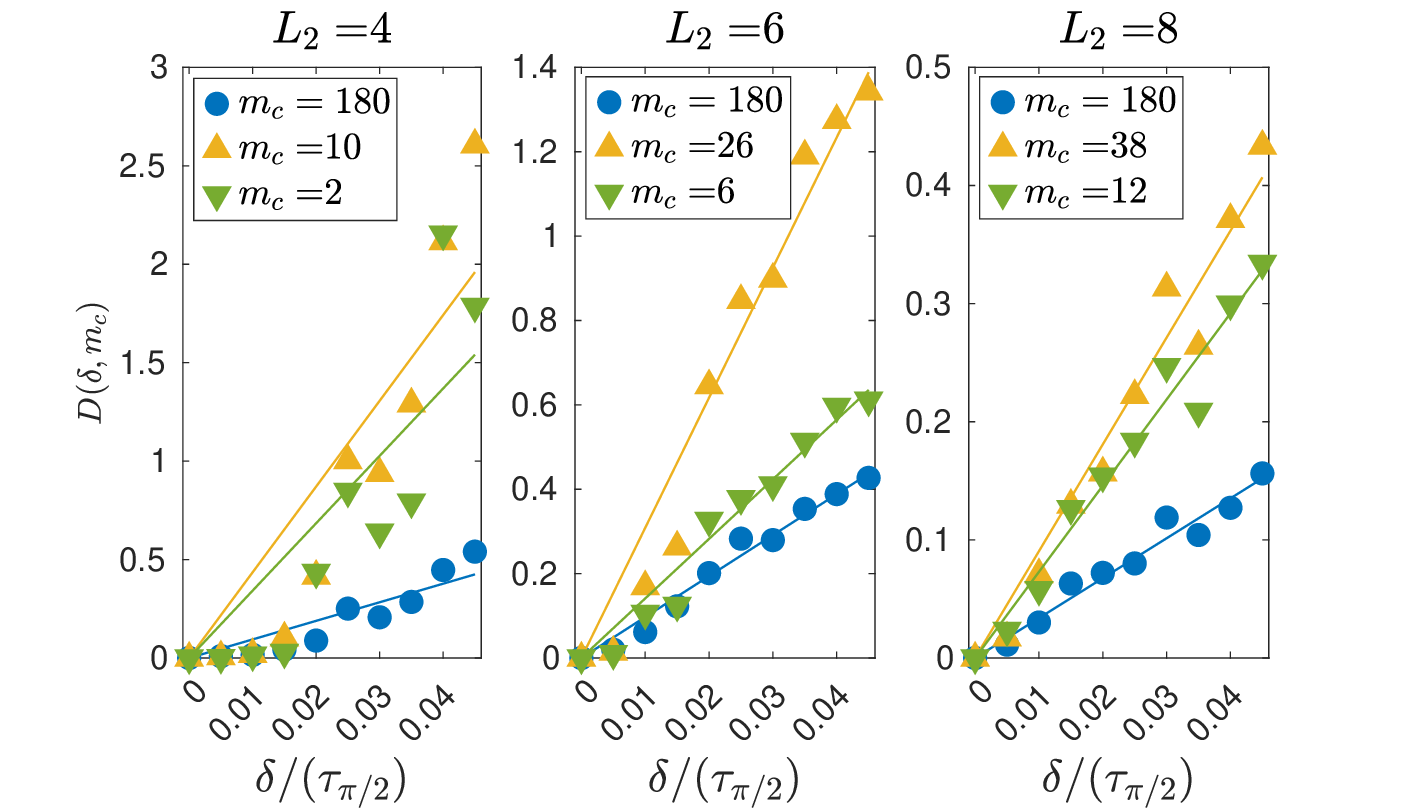}
    \caption{Symbols represent experimental distortion variance $D(\delta,m_c)$  versus pulse width jitter amplitude $\delta$ in units of $\tau_{\pi/2}$ for $L_1=10$  at different $L_2$ values as indicated. Here, the optimal coherence orders are $m_c = 10,~26,~38$.  In each case, two other $m_c$ values are shown for comparision.
    \label{fig:fine}}
\end{figure}

The experimentally measured distortion variance for pulse width jitter amplitudes $\delta \in [0, 0.04\tau_{\pi/2}]$, for cluster building loop number $L_1=10$, and for different sensing durations ($L_2$), with optimal as well as two non-optimal coherence orders, are plotted in Fig. \ref{fig:fine}.  
Firstly, the results show that for $L_2=6$ and $L_2=8$, the distortion variance is systematic, in spite of quantum-correlated clusters of $10^3$ spins involved in sensing the jitters. The linear progression of distortion variance 
supports adaptation for practical sensing via easier calibration and regression.
Interestingly, we find that the optimal coherence orders, although of intermediate values, still yield higher slopes, indicating more efficient sensing than even higher coherence orders.  This highlights the importance of the optimal coherence order in practical quantum sensing.  
For $L_2=8$, the root-mean-square deviation of $D(\delta,38)$ is 0.03, from which we obtain the threshold jitter amplitude as $\sim 10$ ns.  This is a remarkably sensitive detection of the jitter amplitude, given the limitations of the commercial NMR setup, such as over 10\% RF inhomogeneity and other nonlinearities in pulse execution. With only uncorrelated spins, it is hardly practical to achieve such sensitive detection of jitter amplitudes. 



\section{Numerical Modelling}
\label{sec:numerical model}
 In order to gain further insights into the experimental observations, it will be useful to have a numerical model. Full-scale simulation of solid-state experiments with $10^3$ correlated spins is notoriously difficult \cite{hodgkinson2000numerical,tovsner2014computer}.
 Nevertheless, it is illuminating to study a basic model that captures the key features.  We consider permutation invariant states of 40 spins with coherence order distribution resembling that of the experimental spin clusters (see Appendix \ref{appendix:B}). The fact that all experimental readout observables lie in this permutation symmetric subspace justifies this feature of the simulations while making the numerical analysis feasible with limited computing resources. The generated clusters were subjected to a global dephasing channel to model the natural decoherence, and quantum Fisher information (QFI) about the dephasing axis was estimated. The dephasing channel $\mathcal{D}$ is given by 
 \begin{equation}
\mathcal{D}\rho =\prod_i \mathcal{D}_i \rho ,
 \end{equation}
 where the superoperator $\mathcal{D}_i \rho$ can be described by the action
 \begin{equation}
 \mathcal{D}_i \rho = E^i_1\rho E_1^{i \dagger} +E_2^i\rho E_2^{i \dagger}.    
 \end{equation}
 Here, 
 \begin{equation}
 E_1^i = \begin{pmatrix} 1&0\\0&\sqrt{1-p}\end{pmatrix},~
 E_2^i = \begin{pmatrix} 0&0\\0&p\end{pmatrix},   
 \end{equation}
 are the $i^{th}$ spin Kraus operators
 for dephasing strength $p$. 

 \begin{figure}
    \includegraphics[width=0.8\linewidth]{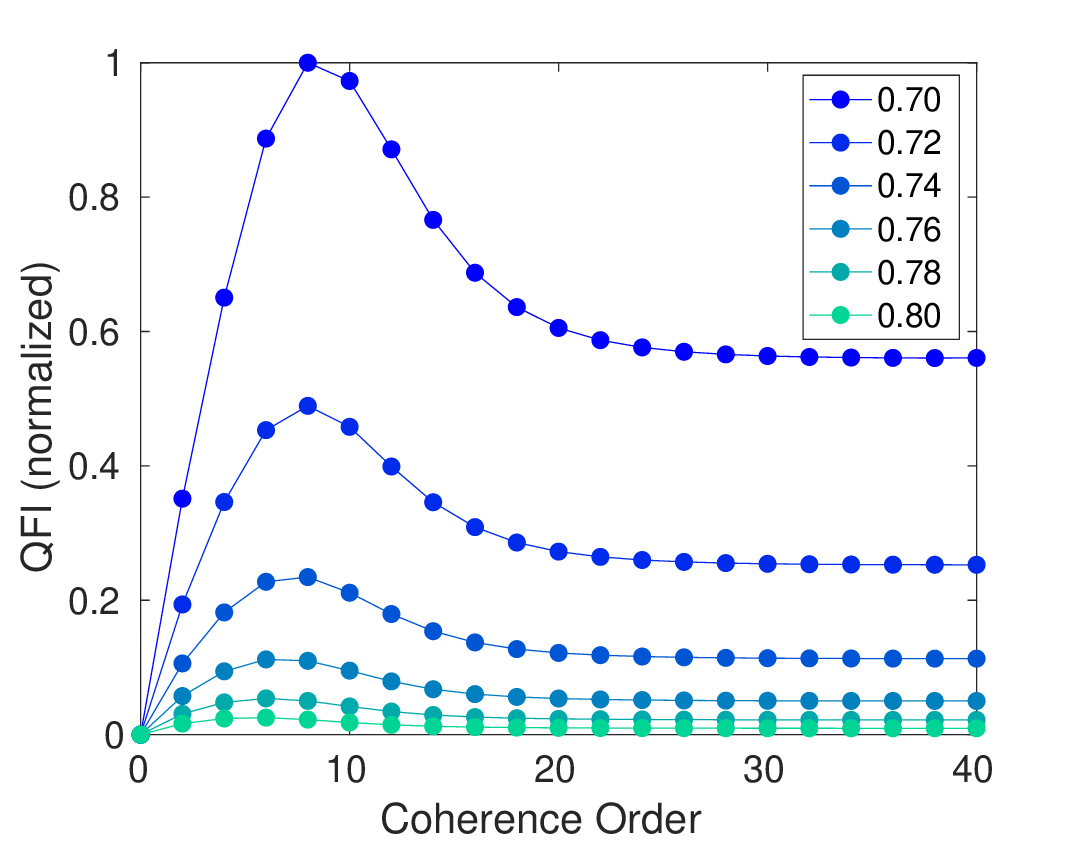}
    \caption{The quantum Fisher information (QFI) versus maximum coherence order $m_c$ for a cluster of 40 spins with dephasing strength $p$ ranging from 0.7 to 0.8.   
\label{fig:fishersimulationhigh}}
\end{figure}

 Using the above numerical model, we estimate QFI of 40 spins for clusters with a range of coherence orders up to $m_c$ to closely follow the experimental scenario. 
 Fig. \ref{fig:fishersimulationhigh}
 plots QFI for a cluster versus $m_c$, with different degrees of dephasing. 
 As expected, we observe an optimum $m_c$ for which QFI peaks.  Of course, it is not straightforward to exactly relate the experimentally measured distortion variance with the numerically estimated QFI. However, the similarity of their trends and the existence of the optimal coherence orders in both cases supports our interpretation of the experimental results and justifies the usefulness of the  numerical model.

 \section{Discussions and conclusions
 \label{sec:Conclusions}}
In this work, we have investigated the applicability of multiple-quantum solid-state NMR for quantum sensing. We first note that the natural Hamiltonian of dipolar-coupled spin clusters can be readily leveraged to generate large correlated states involving hundreds or even thousands of spins. Using a multiple-pulse RF sequence applied to a plastic-crystal sample, we experimentally realize coherence orders of about 100, corresponding to correlated spin clusters of up to ~3000 spins, and extract their coherence-order distributions via phase modulation technique. 

Since quantum entanglement is an important resource for quantum sensing, it is natural to ask whether a solid-state NMR register can be used to probe concepts in many-body quantum metrology.
To explore this, we introduce random RF pulse-width jitters to the large correlated spin cluster and monitor the resulting fluctuations in the coherence-order distributions. 
We define a metric, the distortion variance, to quantify these fluctuations, which in turn serves as a measure for sensing the jitter amplitude. 

A central question is how the sensing efficiency scales with coherence order, given that higher-order multiple-quantum coherences have shorter lifetimes.
To address this, we systematically analyze how the distortion variance depends on the maximum coherence order.  We observe that the sensing efficiency increases up to a certain optimal coherence order and decreases thereafter. This behavior reflects an interplay between the intrinsic sensing capability of each coherence order, its fractional weight in the overall coherence distribution, and its susceptibility to decoherence. Thus, contrary to naive expectations, it is advantageous to exclude the highest coherence orders above an optimal threshold. Using this optimal coherence order, we experimentally detect pulse-width jitters as small as 10 nanoseconds, even in the presence of experimental imperfections such as RF field inhomogeneity.

To gain further insight, we develop a simplified numerical model that emulates key features of the experimental register and then estimate the quantum Fisher information for different coherence orders under dephasing noise. Remarkably, even this minimal model exhibits a similar optimal-order behavior.

We believe that the ability of solid-state NMR to generate, sustain, and manipulate large correlated spin clusters remains underexplored as a resource for quantum sensing. We hope that the present work stimulates further investigations into exploiting these registers for many-body quantum metrology.

 \section*{Acknowledgements}
Authors acknowledge valuable discussions with Vivek Sabarad, Dr. Sandeep Mishra, and Nitin Dalvi.  We thank the National Mission on Interdisciplinary Cyber-Physical Systems for funding from the DST, Government of India, through the I-HUB Quantum Technology Foundation, IISER-Pune.


\appendix
\section{Derivation of master equation}
\label{sec:dissipative}
In order to derive the dissipative superoperator, we first use the toggling frame analysis \cite{avghamiltonian} to obtain the effective average Hamiltonian for a single instance of error values in the sequence shown in Fig. \ref{fig:pulse-sequence}.
We denote the error in the $k$\textsuperscript{th} pulse by $\epsilon_k$. Assuming  $\epsilon_k$ are small, the toggled frame Hamiltonian in each interval is given by:
\begin{equation}
    \begin{split}
        H_1  \approx & H_z \\
        H_2  \approx & H_y + i\epsilon_1[I^x, H_y] \\
        H_3  \approx & H_z + i(\epsilon_1+\epsilon_2)[I^x, H_z]\\
        H_4  \approx & H_y + i(\epsilon_1+\epsilon_2+\epsilon_3)[I^x, H_y] \\
        H_5  \approx & H_z + i(\epsilon_1+\epsilon_2+\epsilon_3+\epsilon_4)[I^x, H_z]\\
        H_6  \approx & H_y + i(\epsilon_1+\epsilon_2+\epsilon_3+\epsilon_4-\epsilon_5)[I^x, H_y] \\
        H_7  \approx & H_z + i(\epsilon_1+\epsilon_2+\epsilon_3+\epsilon_4-\epsilon_5-\epsilon_6)[I^x, H_z]\\
        H_8  \approx & H_y + i(\epsilon_1+\epsilon_2+\epsilon_3+\epsilon_4-\epsilon_5-\epsilon_6-\epsilon_7)[I^x, H_y] \\
        H_9  \approx & H_z + i(\epsilon_1+\epsilon_2+\epsilon_3+\epsilon_4-\epsilon_5-\epsilon_6-\epsilon_7-\epsilon_8)\\
        &[I^x, H_z],
    \end{split}
\end{equation}
where $I^x = \sum\limits_i I^x_i$ and 
$$ H_{y/z} = \sum\limits_{i,j} d'_{ij} \hspace{0.2cm} I^{y/z}_iI^{y/z}_j - \frac{1}{2}(I^{z/y}_iI^{z/y}_j+I^{x}_iI^{x}_j). $$
 Performing the average:
\begin{equation*}
    \begin{split}
        \mathcal{H}_{eff}'  = &\frac{1}{12}[(1/2)H_1+2H_2+H_3+2H_4+H_5\\
          &+2H_6+H_7+2H_8+(1/2)H_9],    
    \end{split}
\end{equation*}
we split $\mathcal{H}_{eff}'=\mathcal{H}_{eff}+\mathcal{H}_{err}$ where
\begin{equation}
\label{eqn:Herr}
    \begin{split}
        \mathcal{H}_{err} = (8\epsilon_1+6\epsilon_2+6\epsilon_3+4\epsilon_4-4\epsilon_5-4\epsilon_6-2\epsilon_7)[I^x, H_y]\\+\frac{1}{2}(7\epsilon_1+7\epsilon_2+5\epsilon_3+5\epsilon_4-3\epsilon_5-3\epsilon_6-\epsilon_7-\epsilon_8)
        [I^x, H_z].
    \end{split}
\end{equation}
We substitute
\begin{equation}
   V= -i[I^x, H_z]  = i[I^x, H_y] = \frac{3}{2}\sum\limits_{i,j} d'_{ij} \hspace{0.1cm} (I^{y}_iI^{z}_j+I^{z}_iI^{y}_j)
\end{equation}
in Eq. \ref{eqn:Herr} to get $\mathcal{H}_{err}$ for a single sampling of $\epsilon_k$.
To get the master equation we begin with the Nakajima-Zwanzig form of the time evolution equation \cite{breuer2002theory}. We treat $\mathcal{H}_{err}$ as a perturbation and move into the interaction picture with respect to $\mathcal{H}_{eff}$. We obtain
\begin{equation}
\label{eqn:nakajima}
    \begin{split}
        d\tilde{\rho}(t)/dt =-i[\tilde{\mathcal{H}(t)}, \tilde{\rho}(0)] - \int_0^t [\tilde{\mathcal{H}(t)},[\tilde{\mathcal{H}(s)},\tilde{\rho}(s)]]ds
    \end{split}
\end{equation}
where $\tilde{A}(t) = e^{i\mathcal{H}_{eff}t}A(t)e^{-i\mathcal{H}_{eff}t}$. Eq. \ref{eqn:nakajima} satisfies the usual Born-Markov and secular approximation since the $\epsilon_k$ are uncorrelated and $\delta$ is small. Expanding out the double commutator and reverting from the interaction picture we get the form in Eq. \ref{eqn:master}.

\section{Numerical Simulation Details}
\label{appendix:B}
The simulations were carried out in the Liouville picture allowing the computation to be done purely in the subspace of operators invariant under spin permutations. A basis of this subspace can be given by operators of the form
\begin{equation}   
\frac{1}{k}\sum_{\pi \in S_{N_s}} \sigma_{\pi^{-1}
(1)}\otimes\sigma_{\pi^{-1}(2)}...\sigma_{\pi^{-1}(N_s)},
\end{equation}
where $k$ is a normalization constant, $\sigma_i$ denotes an operator on the $i^{th}$ spin and $S_{N_s}$ is the symmetric and $N_s$ is the number of spins. The elements of this basis can be uniquely and completely labelled by 4 parameters $N(\sigma^+),N(\sigma^-),N(|0\rangle\langle0|)$ and $ N(|1\rangle\langle1|)$, i.e. the number of terms of each basis element of the single qubit operator space. For convenience however we shall relabel these to an equivalent set
\begin{align*}
    m &=  N(\sigma^+) + N(|0\rangle\langle0|) \\ 
    n &= N(\sigma^-) + N(|0\rangle\langle0|) \\ 
    h &= N(\sigma^+) + N(\sigma^-) \\
    N_s &= N(|0\rangle\langle0|) +N(|1\rangle\langle1|) +N(\sigma^+) + N(\sigma^-)
\end{align*}
for a fixed number of spins we can omit $N_s$ and label each operator as $T^h_{m,n}$ where $m,n$ can go from 0 to $N_s$ and $h$ can go from $\abs{m-n}$ to $\min(m+n,2N_s-m-n)$. In this basis, operator multiplication is given by 
\begin{align}
\label{a1}
T^{h_1}_{m_1,n_1}T^{h_2}_{m_2,n_2} = \delta_{n_1,m_2}\sum_h \chi^h(m1,n1,n2,h1,h2)T^{h}_{m_1,n_2},
\end{align}
where the $\chi^h$ are structure constants that can be computed theoretically. Working in the Liouville picture we represent any permutation symmetric density matrix in this basis:
$$| \rho \rangle\rangle = \sum_{m,n,h} c(m,n,h)|T^h_{m.n}\rangle\rangle$$
Matrices corresponding to the action of all operations and channels on any $| \rho \rangle\rangle$ can be constructed by projecting from the computational basis onto the symmetric subspace basis $T^h_{m,n}$ and using Eq. \ref{a1}. The matrix corresponding to  the dephasing channel $\mathcal{D}$ with strength p takes on a particularly simple diagonal form
\begin{align}
\label{a2}
    \mathcal{D} = \sum_{m,n,h} (\sqrt{1-p})^h |T^h_{m.n}\rangle\rangle\langle\langle T^h_{m.n}|.
\end{align}

\begin{figure}
    \centering
\includegraphics[width=0.8\linewidth]{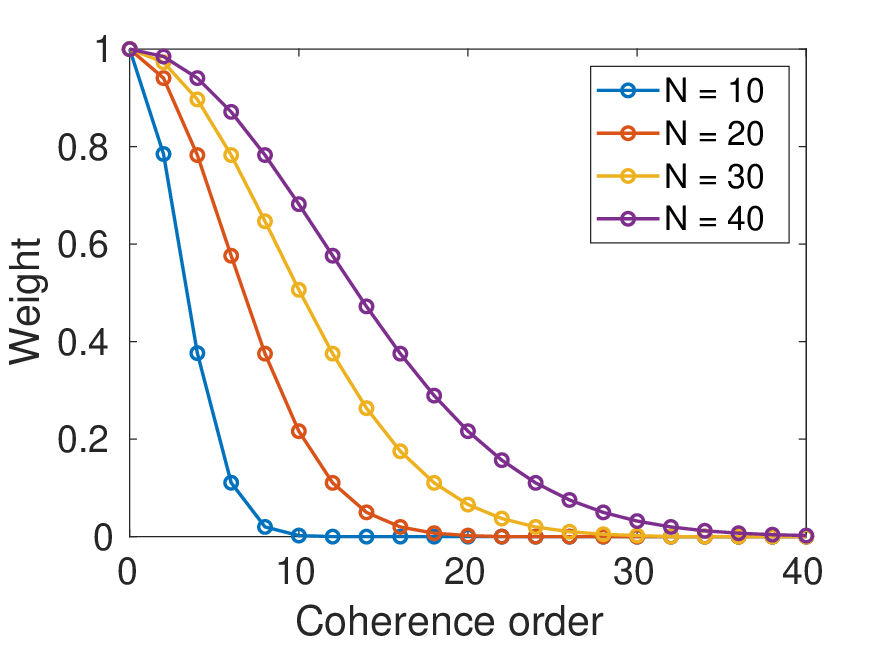}

    \caption{ Normalized coherence order distributions for each total spin value $N$. The width of the Gaussian was chosen to be $N/3.5$ to maximize the width of the Gaussian for that $N$ while simultaneously keeping its value negligible beyond $N$ to prevent spillover. \label{fig:cospectrum}}
    
\end{figure}

For the generation of the cluster states the purely stochastic heuristic outlined in \cite{baum} was used as a guiding principle. Implementing this in the most straightforward way implies that for a cluster with a maximum coherence order $co_{max}$ all density matrix elements $|a_1\rangle\langle a_2|$ (in the computational basis) with even coherence orders from 0 to $co_{max}$ were given an equal weight. After adding the complex conjugate, to make it Hermitian, positivity of the state was ensured by giving the appropriate weight to the corresponding diagonal elements. The resulting distribution of various coherence orders is shown in Fig. \ref{fig:cospectrum}. 

This family of clusters (translated to the Liouville symmetric subspace) was then subjected to the dephasing matrix Eq. \ref{a2}. For calculating the Fisher information in this basis, we used the expression given in \cite{vsafranek2018simple},
\begin{align}
   F_Q= 2 \langle\langle -i [I_z, \hat{\rho}]|(\hat{\rho}^T\otimes \mathbb{1}+\mathbb{1}\otimes\hat{\rho} )^{-1}|-i [I_z, \hat{\rho}]\rangle\rangle,
\end{align}
where $I_z$ is the total z angular momentum operator and $A^{-1}$ denotes the Moore-Penrose  pseudo-inverse of $A$.

\bibliographystyle{ieeetr}
\bibliography{bibliography.bib}
\end{document}